# Transmission of market orders through communication line with relativistic delay

P. B. Lerner[1]

**Abstract**

The notion of "relativistic finance" became ingrained in public imagination and has been asserted in many mass-media reports. Yet, despite an observed drive of the most reputable Wall Street firms to establish their servers ever closer to the trading hubs, there is surprisingly little "hard" information related to relativistic delay of the trading orders. In this paper, the author uses modified M/M/G queue theory to describe propagation of the trading signal with finite velocity.

1. **Introduction**

High-frequency trading (HFT) became feasible and popular the advent of cheap, easily relocated computing power and memory. The times when humans could intervene in the execution of most of the market quotes has long gone. Because of that, a reaction to the changing quotes has to be performed by another computer algorithm, which might react to "stale" prices. (Angel I. J., 2012) The opportunity provided by the modern technology leads to faster and faster trading until time of reaction of electronics (now, in nanosecond range) became small with respect to the relativistic time of propagation of signal between major financial centers (New York-Chicago, 3.8 ms, New York-London, 18.7 ms, New York-Tokyo, 36.2 ms). The characteristic time of the first response to a trading signal is $\tau \approx$ 2-3 ms, which roughly corresponds to the computer messages cycling the circumference of New York City and vicinity with the speed of light, (Hasbrouck, 2016). Inherent latency of trading quotes is even shorter, see, e.g. (Bartlett III, 2019), (see Table 1, *op. cit.*).

The technical advantage of speed was always exploited by the traders. While the case of London's Rothschild receiving detailed information about Napoleon's movements on the continent can be anecdotal, the use of postal pigeons existed since antiquity. (Reuters, 2007) organized their own service across the English Channel since mid-19th century. The first electromechanical fax communicated stock quotes between Lyon and Paris in the 1860s installed by physicist and a priest (Caselli, 1865) but it was highly impractical because of contemporary limitations on technology and was soon replaced by sending coded messages through telegraph. The arms race for the execution time continues to this day. Order execution became fast—time stamps on the order of minutes and seconds were common in early 21st century—but now have reached the microsecond range, for which relativistic limitations on signal transmission became essential (Angel I. J., 2012), (Ang15). To keep up with the progress one has to find a method of analysis, which is largely independent on the extant technology, i.e. practically, of the market latency and trading algorithms.

---

[1] Independent researcher, pblerner@syr.edu

Major exchanges responded to this acceleration with the time stamps up to a fraction of millisecond. (Angel I. J., 2012), (Lewis, 2015)) Since the publication of these references, microsecond stamping became the norm. Geographically separated markets can observe substantially different best executable prices, if the distance between markets $\Delta x > c \cdot \Delta t$, where c is the speed of light and $\Delta t$—the difference between the time order reached the trading venue and the location of the trader. Yet, there are, as of now, no universally accepted financial equations, similar to Black-Scholes, nor pricing models, which explicitly include speed of transmission or node delay.

Namely, in response to the price signal, either external, or endogenous—being generated by an executing algorithm, the terminal (node) of the network generates an order. This order can be canceled on the way if there is opposite order of a larger size; otherwise it propagate further down the network. Our model is designed to quantify imperfections introduced into quote propagation by the finite speed of light and delay introduced by the reaction of other nodes.

The analytic framework proposed by the author is based on modified Takács' notion of a waiting time (Takacs, 1955). We use a modified M/M/G queue theory to describe propagation of the trading signal with finite velocity (Riordan, 1962). In particular, one can establish velocity-dependent analytical condition for market clearing. As a result, we obtain, in the simplest case of a 1 share quote submission, a system of two integro-differential equations, which can be solved part-analytically, part-numerically. This system explicitly contains speed of signal propagation, as well as delay time in a trading network.

The notion of "relativistic finance" fascinates the public imagination and has been asserted in many mass-media reports (Lewis, 2015). Yet, despite an observed drive of the most reputable Wall Street firms to establish their servers ever closer to the trading hubs, "hard" information related to relativistic delay of the trading orders is surprisingly scarce.

2. **M/M/G Transmission Theory**

The state of the server at a time *t* posited at a distance *x* from the trading server at *x=0* (for brevity, we shall call receiving point the "node x" or simply "the node") can be expressed through a virtual waiting time as:

$$p(t,x) \equiv p(w(t) \leq x) \qquad (1)$$

It is equal to the sum of the independent probabilities that the quote does not arrive before the time *t*, $p_1$ and the probability that the trading system reacts during the time between *t* and *t+dt*:

$$p = p_1 + p_2 \qquad (2)$$

The probability $p_1$ can be expressed through *p(t, x)* as follows:

$$p_1 = (1 - a(t)dt)p(t,x) \qquad (3)$$

In the Equation (3), *a(t)* is the arrival rate of the signal, or the intensity of the underlying trading process. It is natural to consider this process as inhomogeneous Poisson or Cox point process, but, for analytic tractability, we shall restrict ourselves to a homogenous Poisson with intensity *a*=const below. The probability $p_2$ is expressed through the response function of the trading system *B(x)*:

$$p_2 = a(t) \int_0^x B(x-y) d_y p(t,y) \qquad (4)$$

The integro-differential equation of Takacš can be written through expanding of the probability of the propagation between quote updates[2] at the time *t+dt* in Taylor series:

---

[2] Note, that Equation (5) is for free propagation—and does not contain Ito terms (Jeanblanc, 2003). Ito equation would appear if one considers a slowly varying response in Equation (6), which can be expanded in Taylor series as well.

$$p(t, x + vdt) = p(t, x) + \frac{\partial p(t,x)}{\partial t} vdt + o(dt) \qquad (5)$$

In (5), the parameter *v* is the velocity of propagation of the electromagnetic signal through the line. Collecting terms from Equations (3), (4) and (5) we obtain:

$$\frac{\partial p(t,x)}{\partial t} = v \frac{\partial p(t,x)}{\partial x} - a(t)p + a(t) \int_0^x B(x-y) d_y p(t,y) \qquad (6)$$

The difference of our situation with the Equation (6) from original Takacs' theory, is that we have <u>two</u> probabilities—the probability of the "Buy" signal and the probability of the "Sell" signal. Both equations must look exactly alike, only with their own set of indexes and boundary conditions.

We presume that a positive *p₁(t,x)* (e.g. "buy") and a negative *p₂(t,x)* (e.g. "sell") signals propagate simultaneously along the network. Henceforth, we replace the Equation (6) with the following system of equations. Technically, this is the system of equations for the signed measures (??) but in what follows, we shall consider probability densities continuous functions of their argument and interpret derivatives in the sense of distributions:

$$\frac{\partial p_1(t,x)}{\partial t} = v \frac{\partial p_1(t,x)}{\partial x} - a(t)p_1 + a(t) \int_0^x B_{11}(x-y) d_y p_1(t,y)$$

$$+ a(t) \int_0^x B_{12}(x-y) d_y p_2(t,y)$$

$$\frac{\partial p_2(t,x)}{\partial t} = v \frac{\partial p_2(t,x)}{\partial x} - a(t)p_2 + a(t) \int_0^x B_{21}(x-y) d_y p_1(t,y) + a(t) \int_0^x B_{22}(x-y) d_y p_2(t,y)$$

$$(7)$$

In the system of Equations (7), the coefficients $B_{11}(x)$ and $B_{22}(x)$ refer to the modification of buy and sell signals during propagation and $B_{12}(x), B_{21}(x)$—cancellation of signals by the signal of opposite sign.

The system of Equations (7) is linear. Propagation of a quote for one share or 1,000 shares takes the same time. However, next level of realism would include a margin for the short selling, which can be modeled as a reflecting boundary at some critical value of *p₂(t,y₀)*. This problem will be investigated elsewhere.

Applying the partial Laplace-Stiltjes transform in x-variable:

$$\varphi_{1,2}(t,s) = \int_0^\infty e^{-sx} dp_{1,2}(t,x)$$

we can as in (Riordan, 1962) obtain a system of ordinary differential equations and integrate it numerically, taking the initial conditions into account. For elucidation of the analytical behavior of the solutions, we apply Laplace transform in both *t* and *x*:

$$\varphi_{1,2}(\tau,s) = \iint_0^{+\infty} e^{-sx-\tau t} dp_{1,2}(t,x)$$

The system of integro-differential Equations (7) requires initial and boundary condition for each variable,

$$\begin{cases} p_1(x,0) = f_1(x,0) \\ p_1(x,0) = f_1(x,0) \\ p_1(0,t) = w_1(t) \\ p_2(0,t) = w_2(t) \end{cases} \quad (8)$$

In real life, the boundary conditions of Equations (8) are stochastic, but for now, we shall consider them arbitrary but known functions.

The system (7) acquires the following form:

$$\begin{pmatrix} \tau - s + a - a\beta_{11}(s) & -a\beta_{12}(s) \\ -a\beta_{21}(s) & \tau - s + a - a\beta_{22}(s) \end{pmatrix} \begin{pmatrix} \varphi_1 \\ \varphi_2 \end{pmatrix} = \begin{pmatrix} \varphi_1(s,0) - sw_1^* \\ \varphi_2(s,0) - sw_2^* \end{pmatrix} \quad (9)$$

For simplicity and clarity of the analytical solution, we suppose $\beta_{11} = \beta_{22} = \beta_1$ and $\beta_{12} = \beta_{21} = \beta_2$. This means, quite intuitively, that the transmission coefficients for the buy and sell orders are the same. Transition matrix in Equation (9) is invertible except for the curves τ=τ(s) in Fourier space outlined by the zeroes of its determinant:

$$P(s,\tau) = [s - \tau - a(1 - \beta_1(s) - \beta_2(s))][s - \tau - a(1 - \beta_1(s) + \beta_2(s))] \quad (10)$$

The zeroes of the determinant of Equation (10) describe, in effect, conditions for the market clearing of the speed-of-light limited trading system. The curves defined by the Equation

$$P(s,\tau(s)) = 0$$

represent the boundaries for the market clearing, which can be interpreted in a language of signal processing. For the waves with the wavenumber $k>s$, the transmission (trading) system acts as a low-pass? Filter with characteristic frequency $\tau(s)$. For the waves with the wavenumber $k<s$, the transmission system acts as a high-pass filter with the same characteristic frequency. These considerations will be presented in the next section in more physical terms of "sending node" and "receiving node", and "past" and "future" (see Fig. 2).

For another simplification, we make a reasonable assumption that the node delay is exponentially distributed. This intuitively corresponds to the fact that traders' quotes are being governed by Poisson random process along the length of the communication line:

$$B_{11,12,21,22}(x) \propto e^{-\lambda|x|}$$
$$\beta_{1,2}(s) \propto \frac{\widetilde{\beta}_{1,2}}{s+\lambda} \quad (11)$$

Where, $\widetilde{\beta}_{1,2}$ are simply c-numbers. Then, one can reduce zeroes of Equation (10) to the zeroes of the characteristic polynomial:

$$P^*(s,\tau) = (s+\lambda)^2[s-\tau-a(1-\beta_1(s)-\beta_2(s))][s-\tau-a(1-\beta_1(s)+\beta_2(s))]$$

The impulse response function (the Laplace transform of the Green function) for the Equation (10) can be found by an inverse Laplace transform. The Laplace original has the form:

$$\widehat{P}^{-1}(s,\tau) = (P^*)^{-1}(s+\lambda)^2 \widehat{M}(s) \quad (12)$$

In Equation (12) 2×2 matrix $\widehat{M}$ does not have singularities outside of the singularities of its denominator (Riordan, 1962), (Takacs, 1955).

With the approximation of Equation (11), inverse of the response function of Equation (10) can be found analytically:

$$\widehat{K}(x,t) \propto \iint_{C_1,C_2} e^{sx+\tau t} \widehat{P}^{-1}(s,\tau)\, ds\, d\tau \quad (13)$$

where $C_{1,2}$ are the contours involved in the inverse Laplace transform (Jeanblanc, 2003), (Davies, 2002). The reaction of the system on an arbitrary trading signal $\vec{\Phi} = \begin{pmatrix} \varphi_1(x) \\ \varphi_2(x) \end{pmatrix}$ can then be computed as the convolution of the matrix defined by Equation (12) as:

$$\vec{F}(x,t) = \int_0^\infty \widehat{K}(x-x',t)\vec{\Phi}(x',0)dx' \qquad (14)$$

Obviously, we do not know anything about the trading signals, which change every millisecond, but in practical applications, one can compute correlation functions of an arbitrary signal using the Equation (14). The time response to any signal, however, is determined by the kernel, $\widehat{K}$.

3. **Representation of the propagating signal**

After simplifications outlined in Section 2, an impulse response (Green) function can be computed semi-analytically. There are several complications involved in visualization of the results.

First, to visualize the Green function, we must accept a convention that is somewhat different from the one familiar from elementary physics textbooks. In most non-cosmological physics contexts, it is natural to imagine a source located at *x=0*, *t=0* and the signal propagating into the future. In the case of trading quotes, we imagine a signal propagating from a distant (but not infinitely distant) past to an observer in the present (see Fig. 2). Indeed, any finite-energy signal propagating from x=-∞ in an absorbing media will have zero amplitude at the point of origin. Thus, we must impose some finite boundary condition on the propagating wave.

Second, and more important complication is that numerical inverse Laplace transform pays no heed to the need to change the path of integration when the argument of the Laplace integrand crosses the Stokes line (Berry, 1988), (Meyer, 1989). This feature has to be entered by hand into a Mathematica© plot. Intuitively, it corresponds to the situation when the signal is completely absorbed at *x=0* and then reemitted by a secondary source in both forward and backward directions.

In Figure 3 we plot the shape of the transfer function near the origin. We observe, naturally, that there is no signal at *x=0* before the origin falls into a future cone for the incoming signal. Propagating signal is asymmetric with respect to the back and front wings and undergoes significant distortion when it penetrates beyond an observation point.

From out simulations we can infer that the maximum distortion happens close to the cusp of a light cone and depends on the bandwidth of the correlation signal only weakly. This happens because close to the cusp of the cone there is no physical difference between forward and backpropagating wavepackets. Only after some time, related to the time of the network reaction (β-coefficients above), there is a distinct separation between transmitted and reflected signal. This conclusion suggests that, in building of the remote trading system, one should not strive to the fastest possible reaction time but select the reaction time of the system, which significantly exceeds the additional latency of the trading signal caused by a finite velocity of light propagation. Otherwise, a response to a delayed signal will be close to random.

We have to establish a certain criterion for fidelity of the transmission of a signal. Because both the baseline distribution and the distribution of an actual transmitted signal are non-stationary, the choice of fidelity measure is far from unique. Fidelity can refer to an initial waveform, as well as to the response of the transmitting line to a delta-shaped distribution. We plot autocorrelation of the Green function as a function of time delay in Fig. 4 and Kullback-Leibler distance $D_{KL}$ between propagating waveforms, interpreted as probability distributions dependent on the time delay at the origin (Fig. 5) (Kullback, 1951), (Burnham, 2002) ,. In our case, we express delay-dependent Kullback-Leibler distance as

$$D_{KL}(\tau) = \int_{-\infty}^{\infty} K\left(x, \frac{x}{c} - \tau\right) \log\left(K\left(x, \frac{x}{c} - \tau\right) \Big/ K\left(x, \frac{x}{c}\right)\right) dx \qquad (15)$$

From the Figure 5, we observe that signal with "maximum information" appears at delay $\tau \approx 2$ in the units of *a* in Equations (9-11) and not at $\tau \approx 0$ as one can imagine. This consideration has to be taken into account in construction of HFT systems, of course, on the basis of much more realistic and network-specific models.

4. **Conclusion**

This paper outlines a theory of signal propagation in the trading network modeled as M/M/G queuing network. Namely, the M/M/G station at x=0 services a signal propagating from the left (theoretically, from -∞). Trading signal is modeled as a signed probability measure according to the specification developed in a pioneering work of (Takacs, 1955). Update of quotes is represented by the exponentially correlated distribution, or a continuous version of Poisson process.

The main conclusion of this theory is that the response of a trading node instead of being as quick as possible, must be harmonized with the parameters of the entire network. The intuitive reason for that is that every trading network is dispersive. Even a white (delta-correlated) noise acquires finite spread during propagation in the network. At the current node located at the apex of the light cone, there is no physical difference between the front of an advanced signal and the tail of a retarded signal. Only when advanced and retarded noise are sufficiently separated in time, can the trading signal be resolved.

**Figures**

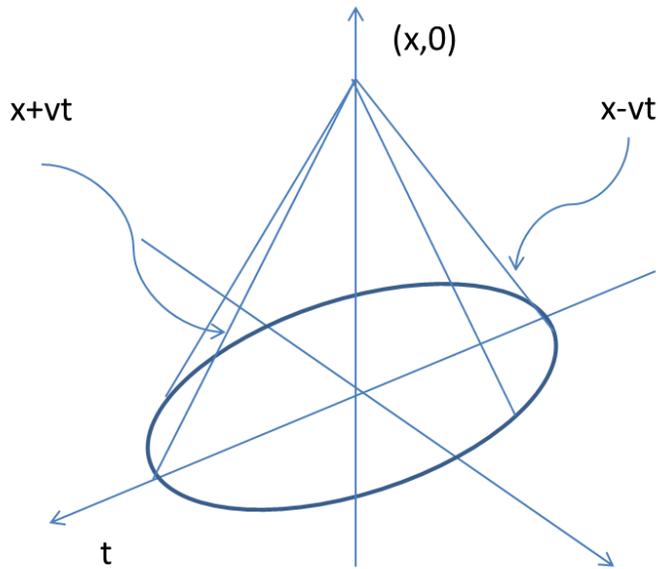

A)

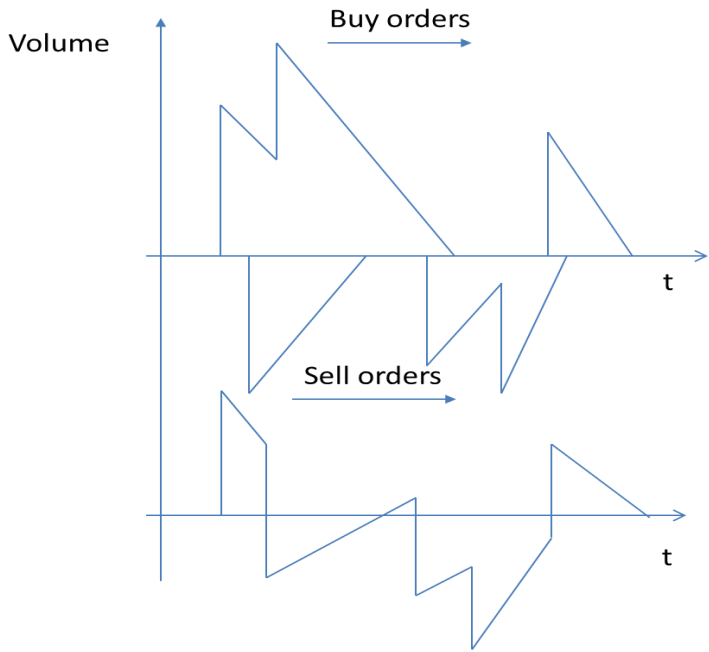

B)

Figure 1. A) Relativistic light cone. Only part x>0 is drawn for clarity. B) Schematic shape of the random signal propagating across the line. Upper axis shows "Buy" and "Sell" orders separately, the lower axis shows sum of the orders.

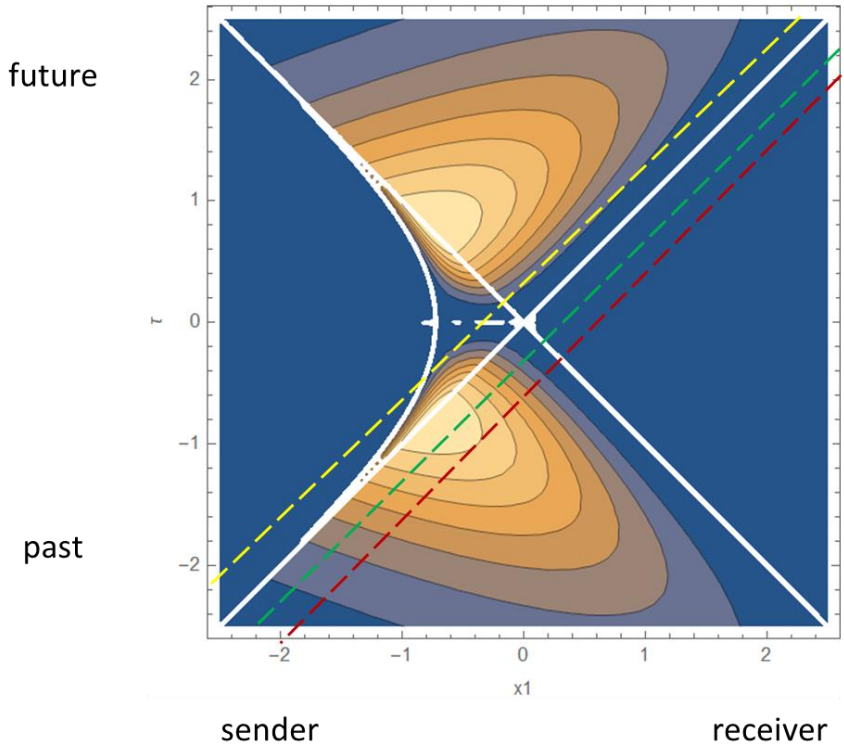

Fig. 2 Spacetime propagation of the signal in the trading system. The purple line indicates $\tau = x/c - 0.5\,a$ (the signal is in the future with respect to observer at the origin) the green line $\tau = x/c - 0.25\,a$ (signal approaches the origin) and the yellow line—$\tau = x/c + 0.25\,a$ (the signal passes the origin). One might consider $a$ as a crude measure of the reaction time of the trading system. Values for the parameters of the Equations (9), (11) and (12) used for all figures are $\lambda = 1.5\,a$, $\tilde{\beta}_{11} = \tilde{\beta}_{22} = 0.6\,a$, $\beta_{12} = \beta_{21} = 0.3\,a$ and were chosen for the best visual appearance of plots.

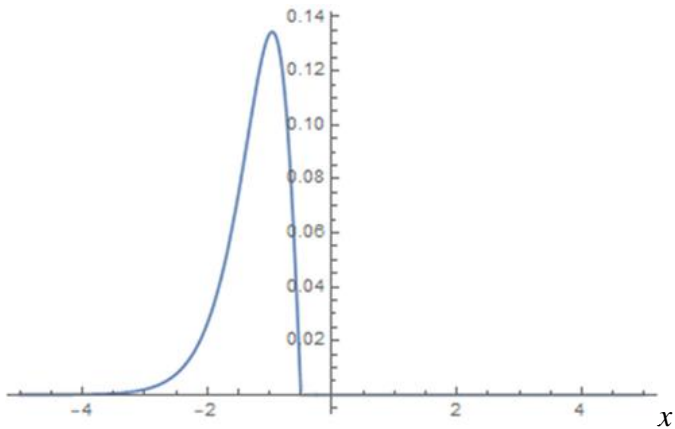

A)

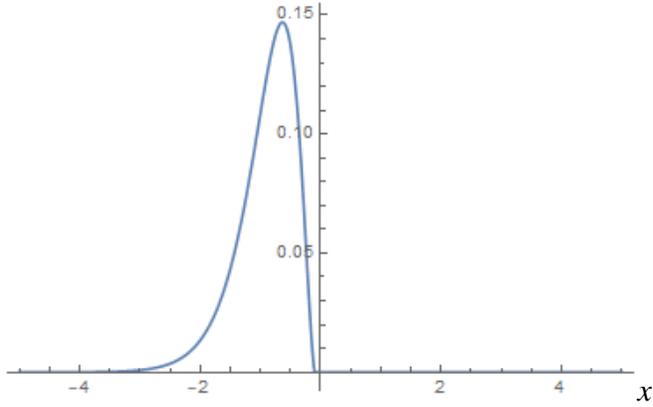

B)

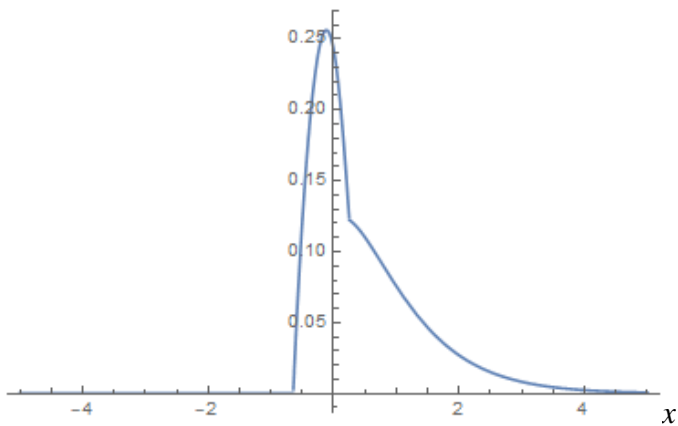

C)

Fig. 3. The shape of the pulse in arbitrary units (a response to the delta function) in the cases A) $\tau = x/c - 0.5\,a$ , B) $\tau = x/c - 0.25\,a$ , C) $\tau = x/c + 0.25\,a$ . The spatial coordinate is measured in the units $c \cdot a$ (see Equation (5)).

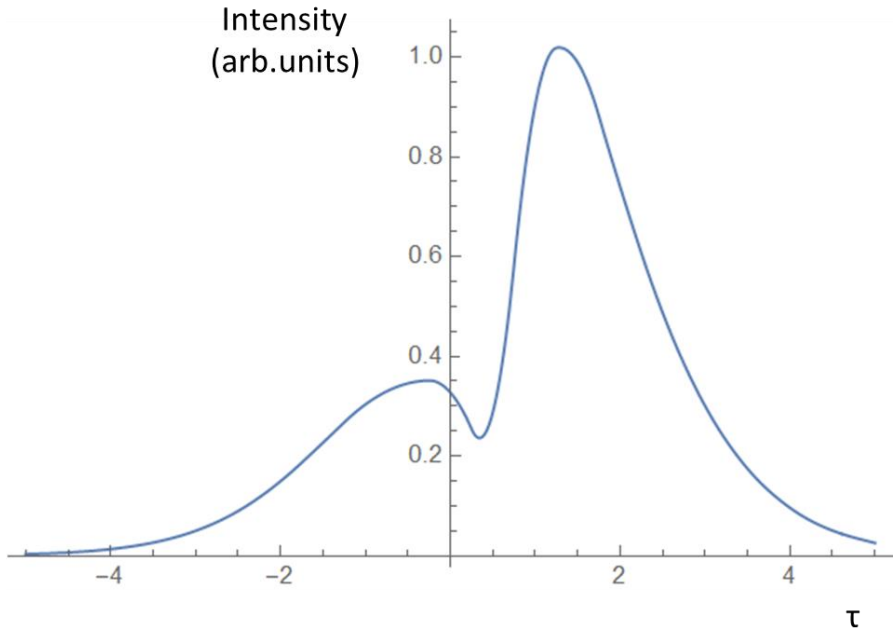

Fig. 4. Autocorrelation of the square of the Green function kernel $I(\tau) = \int_{-10}^{10} K^2(x, \frac{x}{v} - \tau)dx$ (see Equation (14) in the text) as a function of time delay and $v=0.75c$. Limits of integration are arbitrary as a truncation to approximate $[-\infty, +\infty]$ with a necessary accuracy.

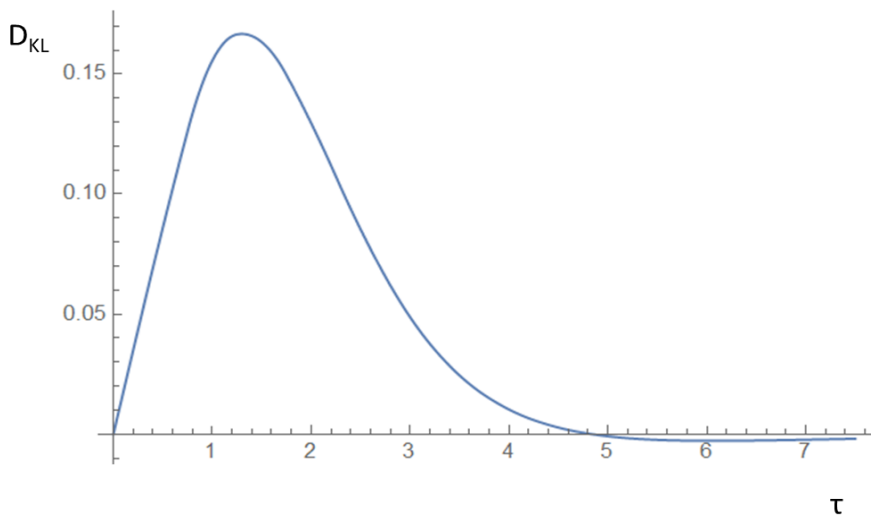

Fig. 5. Kullback-Leibler distance (Equation (15)) between Green functions kernels as a function of time delay and $v=0.75c$. Small negative tail is spurious and is related to numerical approximations.

# References

Angel, 2012. Angel, J. J., L. E. Harris, D. McCabe, 2012, Fairness in Financial Markets: the Case of High Frequency Trading, Journal of Business Ethics, Journal of Business Ethics DOI: 10.1007/s10551-012-1559-0.

Angel, 2015. Angel, J. J., L. E. Harris, C. S. Spatt, Equity Trading in the 21$^{st}$ Century, Quarterly Journal of Finance, vol. 5(1), 1-39.

(Bartlett III, 2019) Bartlett III, R. P. and J. McCrary, How rigged are stock markets? Evidence from microsecond timestamps, Journ. of Financial Markets, 45, 37-60.

(Reuters, 2007), https://www.theguardian.com/media/2007/may/04/reuters.pressandpublishing

Caselli, 1865, Giovanni Caselli, en.wikipedia.org/wiki/Giovanni_Caselli, last downloaded 2/17/2020

(Hasbrouck, 2016) Hasbrouck, J. YouTube FMA Lecture, Dec. 29, 2016. https://www.youtube.com/watch?v=EZCgW1mFRP8&t=1905s+

Takacs, L., 1955, Investigation of waiting time problems by reduction to Markov processes, Acta Math. Acad. Sci. Hung., Vol. 6, pp. 101-129.

Lewis, M., 2015, Flash Boys, W. W. Norton, New York, NY.

(Riordan, 1962) Riordan, J., 1962, Stochastic Service Systems, John Wiley: New York, NY.

(Jeanblanc, 2003) Jeanblanc, M., M. Yor and M. Chesney, 2003, Mathematical Methods for Financial Markets, Springer: New York, NY.

(Davies, 2002) Davies, B., Integral Transforms and Their Applications, 2002, Springer: New York.

(Berry, 1988) Berry M. V., Uniform asymptotic smoothing of Stokes' discontinuities, Proc. Roy. Soc. London, 1989, A422, 7-21.

(Meyer, 1989) Meyer, R. E., A simple explanation of the Stokes phenomenon, SIAM Rev., 1988, 31(3), 435-445.

(Kullback, 1951) Kullback S. and Leibler R. A., On information and sufficiency, Annals of Mathematical Statistics, 1951, 22(1), 79-86.

(Burnham, 2002) Burnham K. P. and D. R. Anderson, Model Selection and Multi-Model Experience (2$^{nd}$ ed.), 2002, Springer: Heidelberg, FRG.